# Miniaturized atmospheric ionization detector

Karen Aplin[1,2,*], Aaron Briggs[1], Adam Baird[1], Peter Hastings[1], R. Giles Harrison[2], Graeme Marlton[2]

1. Department of Physics, University of Oxford, Oxford, UK
2. Department of Meteorology, University of Reading, Reading, UK

**ABSTRACT:** A small scintillator-based detector for atmospheric ionization measurements has been developed, partly in response to a need for better ionization data in the weather-forming regions of the atmosphere and partly with the intention of producing a commercially available device. The device can measure both the count rate and energy of atmospheric ionizing radiation. Here we report results of a test flight over the UK in December 2017 where the detector was flown with two Geiger counters on a meteorological radiosonde. The count rate profile with height was consistent both with the Geigers and with previous work. The energy of incoming ionizing radiation increased substantially with altitude.

**INTRODUCTION**

There is increasing interest in energetic particles in the atmosphere. Ionisation from cosmic rays and, occasionally, space weather events may have small effects on weather and climate [e.g. Mironova et al, 2015]. Despite the associated need for lower atmosphere measurements, our understanding of these particles mainly depends on satellite measurements of primary particles and surface measurements of secondary particles. The world-wide neutron monitor network provides an effective measurement of the cosmic ray nucleonic component [e.g. Simpson, 2000], but this non-ionising radiation is uncorrelated with ionizing radiation in the lower atmosphere [Harrison et al, 2014]. There are few measurements of ionizing radiation in the weather-forming regions of the atmosphere, and whilst models do exist, there are known problems with them due to their reliance on standard thermodynamic profiles [e.g. Usoskin and Kovaltsov, 2006].

These factors have motivated development of a miniature (mass of 30g) scintillator-based detector coupled to a PiN diode, which responds to ionising radiation from gamma and cosmic ray sources [Aplin et al, 2017]. The instrument can be deployed on meteorological radiosondes or unmanned airborne vehicles, or in a stand-alone mode as a hand-held device with data output via Bluetooth or USB. It is sufficiently inexpensive to be disposable. Unlike a conventional Geiger counter, this detector can be run from a low bias voltage (12 V) at a low current (20 mA). Furthermore it has the important advantage of being able to resolve particle energy, which allows identification of different types of ionising radiation. In this paper we describe results from a test flight on 6th December 2017, when the new instrument, here called the PiN detector, was flown on a meteorological radiosonde with two Geiger counters for comparison.

**TEST FLIGHT**

---

* Contact information:   Dr Karen L. Aplin, Department of Physics, University of Oxford, Denys Wilkinson Building, Keble Road, Oxford, OX1 3RH, UK, karen.aplin@physics.ox.ac.uk

*Balloon instrumentation*

The PiN detector generates pulses in response to detected energetic particles. As well as the pulse rate, a pulse size measurement, proportional to the energy, is obtained by differencing the pulse height from the background [Aplin et al, 2017]. In a laboratory context the size of each individual pulse can be readily determined and analysed together with the pulse rate, however data transmission rate limitations for radiosonde use require the pulse size to be determined in situ, with only the count of events of different energies sent over the radio link. For redundancy, the raw reference level and pulse height data are also returned from the last count measured before return of the data packet.

For the test flight, the PiN was interfaced with a Vaisala RS92 radiosonde using the PANDORA system [Harrison et al, 2012], along with two LND614 Geiger counters, responding to beta and gamma radiation. The PANDORA data is additional to the standard meteorological data, and is returned by the radiosonde through a data channel normally used for ozone measurements. The Geiger counters returned the number of counts per timestamped data packet, whereas the PiN sensor sent back the total number of counts, the number of counts in each of eight energy bins, plus the reference level and pulse size from the last pulse recorded. The RS92 radiosonde also measured the standard meteorological quantities of air pressure, temperature, wind and humidity, which are unaffected by the PANDORA system.

*Flight details*

The balloon was launched at 1455UT on 6$^{th}$ December 2017 from Reading University Atmospheric Observatory (51.46° N, 0.95° W), between rain showers. The surface conditions at launch were cloudy, with the cloud base at about 1.2 km, temperature 10°C, and a light south-south-westerly breeze of 2 m/s. Space weather conditions (from spaceweather.com) were calm with a Kp index of 3, an interplanetary magnetic field of 4.5 nT and no sunspots. The balloon burst at 26 km (51.33° N, 0.03° E) and the parachuted package fell to ground, with contact lost at approximately 1800UT near Maidstone in Kent (51.29° N, 0.44° E), 98 km from Reading. Meteorological data measured during the ascent showed that the conditions were similar to those in Reading. The trajectory and basic meteorological results are shown in Figure 1.

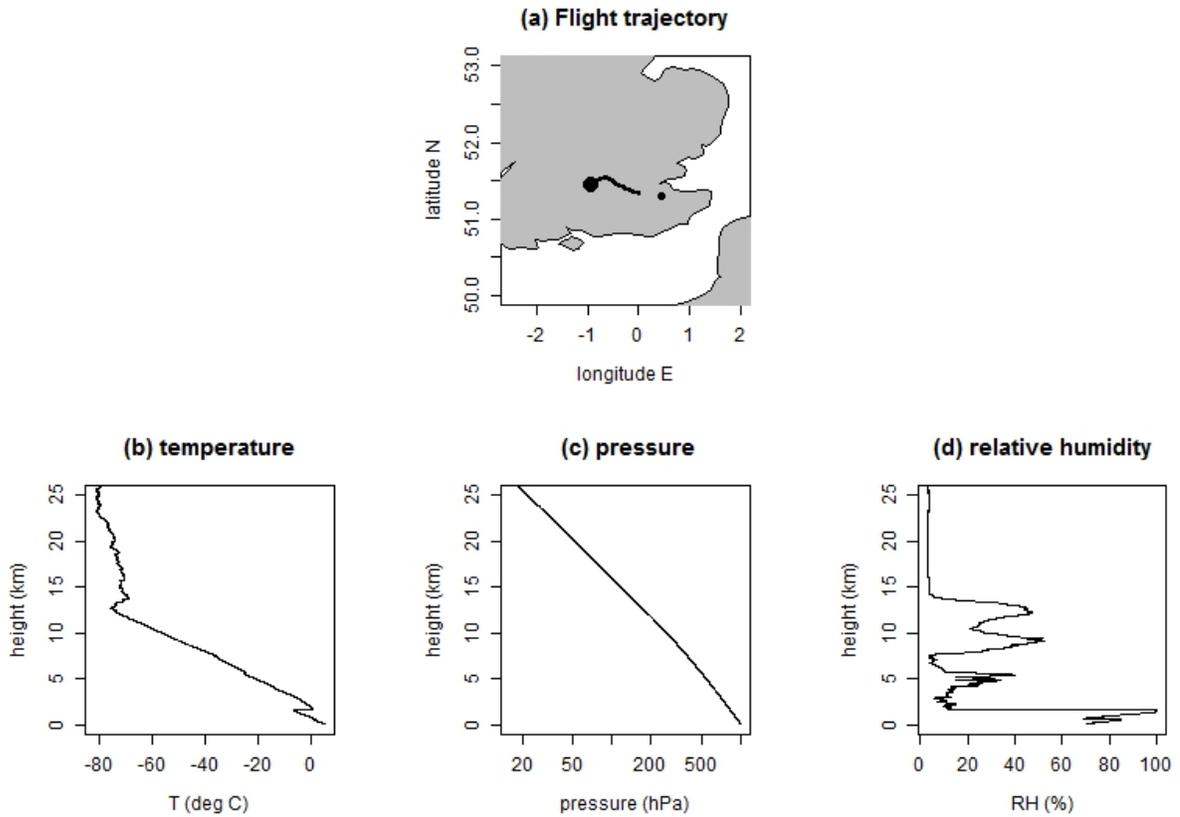

Figure 1 Overview of flight on 6th December 2017 (a) Map showing trajectory from launch site (large black dot) to burst position, with landing position indicated with a small black dot (b) Temperature profile (c) Pressure profile (d) Relative humidity profile

**RESULTS**

*Count rates*

Both the Geiger counters and the PiN detector increased their count rate with height, as expected. The profiles of raw count rates with altitude were consistent in shape between the Geigers and the PiN detector, as can be seen in Figure 2. There were differences between the Geiger counters of more than the 2% reported in Harrison et al [2014] in laboratory tests; Geiger 1 has a clear maximum at ~15km, whereas Geiger 2 and the raw PiN counts do not have a well-defined maximum. The PiN detector's quantum efficiency (the number of pulses returned per gamma ray interaction with the scintillator) decreases with temperature, implying that the increase in count rate shown in Figure 2(b) will be further enhanced, as indicated in the corrected count rate plot Figure 2(c).

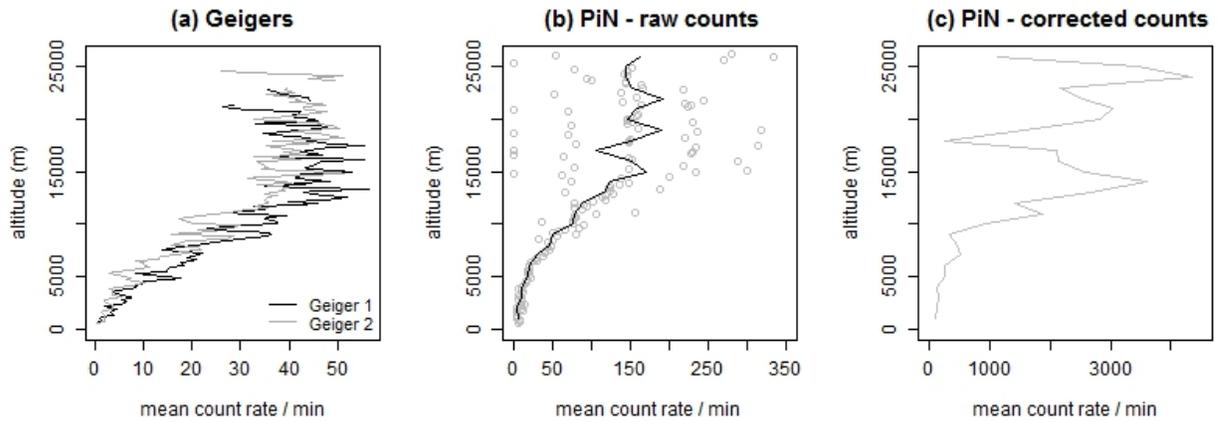

Figure 2 Ionising radiation atmospheric profiles (a) Geiger counters (b) Raw counts from PiN detector with an overplotted median line for 1km altitude bins (c) PiN detector corrected for temperature effects on the sensor's quantum efficiency.

Previous Geiger flights in 2013-2014 from the same site [Harrison et al, 2014] showed a better-defined maximum in count rate with altitude. This maximum in ionization, or count rate, is known as the Regener-Pfotzer maximum after its discoverers who flew both Geiger tubes and ionization chambers on balloons [Carlson and Watson, 2014]. The clarity of the Regener-Pfotzer (RP) maximum is known to vary with the solar cycle for measurements made at a single site. At solar minimum (cosmic ray maximum) the RP maximum is more poorly defined than at solar maximum (cosmic ray minimum) [e.g. Neher and Anderson, 1964]. The Reading measurements in 2013 were near solar maximum, whereas the 2017 data is approaching solar minimum, when the RP maximum is expected to be less well defined. The Geiger counters and PiN results are therefore consistent with previous measurements.

*Energy profiles*

The counts returned were binned by pulse size, corresponding to energy. The number of counts in each bin plotted against altitude indicates a strong increase in energy, Figure 3. Gamma rays in the atmosphere originate both from terrestrial radioactivity of typical energy 0.1 – 2.6 MeV, and from secondary cosmic rays, which have a wide energy range [Baldoncini et al, 2018]. The secondary cosmic rays form part of the electromagnetic cascade, responsible for much of the ionisation in the atmosphere above the troposphere, where muons are the dominant cosmic ray source [Usoskin and Kovaltsov, 2006]. The terrestrial radioactivity contribution is well-known to decrease with distance from the terrestrial surface, as can be seen in the shape of the lowest energy bin, centred on ~200 keV.

Comparing the altitude of the maximum count rate in each energy bin with the Geiger counters, Geiger 1 and the 3 MeV channel both have a maximum at 15km, whereas Geiger 2 and the greatest energy bin (7-19 MeV) both have no clear maximum. This could indicate that the two Geigers have a slightly different energy response, with Geiger 2 responding to more energetic particles than Geiger 1.

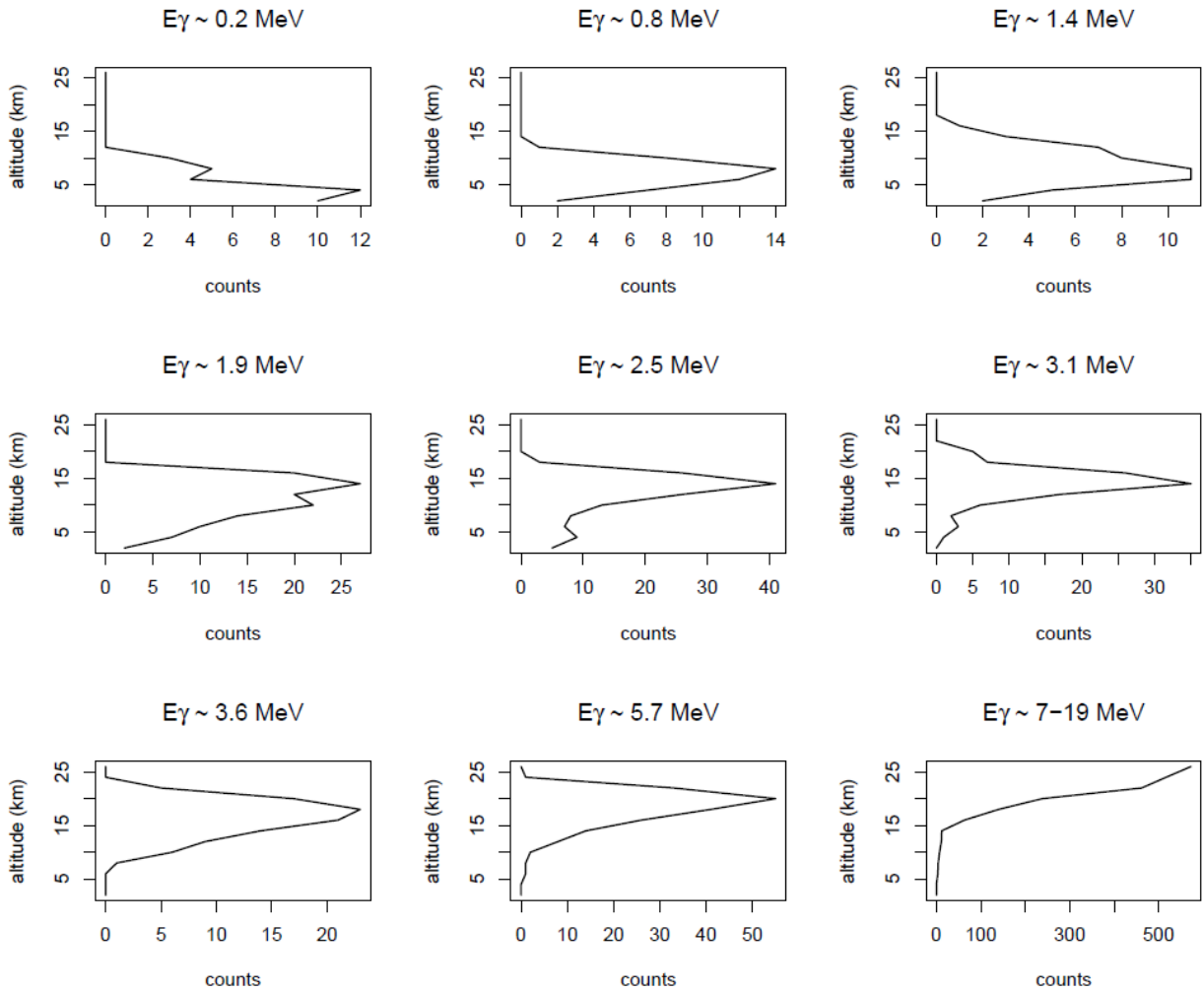

Figure 3. Profiles of counts in each pulse-height bin. The approximate gamma energy at the centre of each bin is indicated in the heading. The data are not corrected for temperature effects on quantum efficiency nor scintillator yield.

**DISCUSSION**

A new type of atmospheric ionization detector, the PiN detector, has been tested on a meteorological balloon flight over the UK in December 2017, alongside two Geiger counters. Meteorological and space weather conditions were calm and relatively consistent over the flight duration and are therefore unlikely to affect the instrument's performance. The count rates obtained were consistent with previous Geiger flights, once variations in the solar cycle have been taken into account.

Particle energies were also measured and found to increase considerably with altitude. The detector has only been calibrated for gamma rays, so caution must be taken in literal interpretation of the energies at higher altitudes where it is likely that the detector is also responding to other ionizing radiation, such as primary cosmic rays (mainly protons and energetic alpha particles), which it has not yet been calibrated for. Despite this, it is quite clear that the number of the greatest energy particles in the atmosphere increases with altitude. This indicates that certain particle energies are dominant in weather-forming regions of the atmosphere, and that ionization from particles in this range of energies may disproportionately affect atmospheric processes over that from particles detected outside the atmosphere, or at the surface. It is found that the Regener-Pfotzer maximum is a poorly-defined quantity that varies with the instrument used, both in terms of the energies of particles being measured, and between nominally identical detectors.

Further work is under way to understand the response of the detector to non-gamma ionizing radiation, to convert count rate to ionization rate, and to fully characterize its efficiency. The response of the detector to energetic particles produced from thunderstorms will also be investigated. Finally, the PiN device is being commercially developed and is available for sale as a stand-alone device or for integration with other instrumentation.

**ACKNOWLEDGMENTS**

This project was funded by the UK Science and Technology Facilities Council Impact Accelerator Account.